\documentclass[preprint,aps,12pt,amsmath,amssymb,showkeys,nofootinbib,tightenlines,superscriptaddress]{revtex4-1}
\usepackage{amssymb}
\usepackage{amsfonts}
\usepackage{latexsym}
\usepackage{pdfsync}
\usepackage[T1]{fontenc} 
\usepackage{epsf,amsfonts}
\usepackage{epsfig}
\usepackage{amsthm}
\usepackage{amsmath}
\usepackage{graphicx}
\usepackage{color}
\usepackage{slashed}
\usepackage{mathrsfs}
\usepackage{float}
\usepackage{color}
\usepackage{esvect}
\usepackage{mathtools}

\newcommand{\VEV}[1]{\left\langle #1\right\rangle}

\newcommand{\MeV}{\;\text{MeV}}

\newcommand{\re}{\mathrm{Re}}

\begin{document}

\title{Chiral phase transition and QCD phase diagram from AdS/QCD}


\author{Zhen Fang}\email{fangzhen@itp.ac.cn}\affiliation{Department of Applied Physics, School of Physics and Electronics, Hunan University, Changsha 410082, China}
\author{Yue-Liang Wu}\email{ylwu@itp.ac.cn}\affiliation{CAS Key Laboratory of Theoretical Physics, Institute of Theoretical Physics, Chinese Academy of Sciences, Beijing 100190, P. R. China}\affiliation{International Centre for Theoretical Physics Asia-Pacific (ICTP-AP), University of Chinese Academy of Sciences, Beijing 100049, China}
\author{Lin Zhang}\email{zhanglin@itp.ac.cn}\affiliation{CAS Key Laboratory of Theoretical Physics, Institute of Theoretical Physics, Chinese Academy of Sciences, Beijing 100190, P. R. China}\affiliation{School of Physical Sciences, University of Chinese Academy of Sciences, Beijing 100049, China}

\date{\today}

\begin{abstract}
We study the chemical potential effects on the chiral phase transition in a simply improved soft-wall AdS/QCD model, which can realize consistently the properties of linear confinement and spontaneous chiral symmetry breaking. The $\mu-T$ phase diagrams with both zero and physical quark masses have been obtained from this model. For the case of physical quark mass, the chiral transition has a crossover behavior at low chemical potential. With the increase of the chemical potential $\mu$, the critical temperature $T_c$ descends towards zero, and the crossover transition turns into a first-order one at some intermediate value of $\mu$, which implies that a critical end point naturally exists in this improved soft-wall model. All these features agree with our current perspective on the QCD phase diagram.
\end{abstract}

\keywords{chiral phase diagram, chemical potential effects, 2+1 flavor, AdS/QCD}

\maketitle
\section{Introduction}\label{introduce1}

The dynamics of quarks and gluons in quantum chromodynmaics (QCD) has significant influence on the evolution of the universe and the existence of our world, which is intimately related to the two important features of QCD, i.e., the color confinement as a result of the nonlinear and nonperturbative dynamics of QCD and the mass generation originating from the quark condensate. At extreme environments such as high temperature or high baryon density, the confined hadronic phase will dissociate into quark-gluon plasma (QGP) with color released and also accompanied with the restoration of chiral symmetry. With the deepening of research, it has been known that the QCD phase structure is far beyond the simple picture stated above. Theoretical understanding of the whole QCD phase diagram and the mechanism for the formation of the rich phase structure is still a challenge in modern physics.

At zero baryon density, the quark flavors and their masses have a significant impact on the properties of QCD phase transition, which are clearly summarized in the Columbia plot \cite{Brown:1990ev,Laermann:2003cv}. The qualitative features of QCD phase transition can be extracted from the universality analysis of effective models when quark masses approach to zero or infinity. In the infinite quark-mass limit, the effective theory for the order parameter, viz. the Polyakov loop expectation value characterizing the behaviour of the heavy quark free energy at large distances, is a three dimensional spin model with global Z(3) symmetry, which signifies a first-order phase transition \cite{Svetitsky:1982gs}. In the vanishing quark-mass limit, the flavor dependence of QCD phase transition has been investigated by a renormalization-group analysis of an effective model for the chiral condensate, which is the proper order parameter in this case \cite{Pisarski:1983ms}. Lattice QCD studies in recent years have shown strong evidence on a crossover transition with the pseudocritical temperature $T_c\simeq 155\MeV$ at physical quark mass, and a second-order one in the two-flavor case with zero quark mass \cite{Aoki:2006we,Bazavov:2011nk,Bhattacharya:2014ara}. A great deal of information for the properties of QCD phase transition have been attained from both lattice and phenomenological studies and also from experimental researches.

At finite baryon density, the QCD phase structure is further enriched significantly \cite{Stephanov:2007fk,Fukushima:2013rx}. Besides the hadronic phase and the QGP phase, there may also exist other phases of strongly interacting matters at large chemical potential $\mu$ and low temperature $T$, such as the quarkyonic phase \cite{McLerran:2007qj} and the color superconducting phase \cite{Rapp:1997zu,Alford:1997zt}. With the increasing of the baryon chemical potential, it is generally believed that the crossover transition will evolve towards a critical end point (CEP) and then turn into a first-order phase transition. However, there remain questions regarding the existence and the location of CEP \cite{Fodor:2001pe,Ejiri:2003dc,Gavai:2004sd,Endrodi:2011gv,Qin:2010nq}. Lattice QCD extrapolation to the finite chemical potential case is rather difficult due to the notorious sign problem and has so far not reached to a consistent conclusion. Hence, other approaches such as the Dyson-Schwinger equation and the functional renormalization-group method \cite{Roberts:2000aa,Fischer:2009wc,Fischer:2011mz,Braun:2006jd, Braun:2009gm} and effective models like the Polyakov-loop extended Nambu-Jona-Lasinio model \cite{Fukushima:2003fw,Ratti:2005jh} have been used to address the relevant issues on QCD phase diagram. Experimentally, there are also many ongoing and planning programs aimed to probe the signals of CEP and other intrinsic properties of QCD phase structure \cite{Sorin:2011zz,Andronov:2015ucu,Lacey:2015yxg,Mustafa:2015yeg, Luo:2015ewa}.

In recent decades, the Anti de-Sitter/conformal field theory (AdS/CFT) correspondence \cite{Maldacena:1997re,Gubser:1998bc,Witten:1998qj} provides another powerful tool to tackle the strongly interacting issues of low-energy QCD, which is usually called AdS/QCD. There are essentially two approaches for the model construction in AdS/QCD, i.e., the top-down approach which set off from string theory and the bottom-up approach which is based on the fundamental properties of low energy QCD. As a promising way to tackle the nonperturbative low-energy problems of strong interaction, this field has attracted extensive studies in the past decades, especially on the issues of hadron spectrum, QCD thermodynamics and many other aspects of low-energy hadron physics \cite{DaRold:2005mxj,Erlich:2005qh,Karch:2006pv,deTeramond:2005su,Brodsky:2014yha,Babington:2003vm, Kruczenski:2003uq,Sakai:2004cn,Sakai:2005yt,Csaki:2006ji,Cherman:2008eh,Gherghetta:2009ac,Kelley:2010mu,Sui:2009xe,Sui:2010ay,Cui:2013xva, Li:2012ay,Li:2013oda,Shuryak:2004cy,Tannenbaum:2006ch,Policastro:2001yc,Cai:2009zv,Cai:2008ph,Sin:2004yx,Shuryak:2005ia,Nastase:2005rp, Nakamura:2006ih,Sin:2006pv,Janik:2005zt, Herzog:2006gh,Li:2014hja,Li:2014dsa,Fang:2016uer,Fang:2016dqm,Fang:2016cnt,Fang:2016nfj, Evans:2016jzo,Mamo:2016xco, Dudal:2016joz,Dudal:2018rki,Ballon-Bayona:2017dvv,Chelabi:2015cwn,Chelabi:2015gpc}. In this work, we focus on the properties of chiral phase transition and QCD phase diagram at finite chemical potential. The relevant issues have been studied in the well-known Sakai-Sugimoto model in the probe approximation \cite{Sakai:2004cn,Horigome:2006xu}, which leads to a very different phase structure from the QCD expectations. Other top-down models such as that using D6-brane or D7-brane probes in some specific dual backgrounds have also been proposed \cite{Babington:2003vm,Kruczenski:2003uq}, and so far there are no satisfying descriptions for chiral phase transition in the top-down approach.

In spite of the clear stringy origins of the top-down approach, it seems that we still need much more to construct a realistic AdS/QCD model. Thus it deserves for us to switch to the bottom-up approach temporarily and to find a possible phenomenological way to realize the expected chiral phase transition and the QCD phase diagram from holography, which indeed have attracted intense works in the bottom-up AdS/QCD. Most of the studies are focused on the issues of deconfining phase transition and the associated thermodynamics \cite{Herzog:2006ra, Andreev:2006eh,Andreev:2009zk,Gubser:2008yx,Gubser:2008ny,Gubser:2008sz,BallonBayona:2007vp,Cai:2007zw,Kim:2007em,Gursoy:2008za, Gursoy:2008bu,Lee:2009bya,Jo:2009xr,Colangelo:2010pe,Li:2011hp,He:2013qq,Yaresko:2013tia,Yang:2014bqa,Cui:2014oba,Finazzo:2014zga,Fang:2015ytf, Critelli:2017oub,Rougemont:2015wca,Knaute:2017opk}. To address the issue of chiral phase transition with finite chemical potential, we follow the original line of bottom-up approach, i.e., the hard-wall and soft-wall models \cite{DaRold:2005mxj,Erlich:2005qh,Karch:2006pv}. We also note that another researching line with more stringy ingredients included has been given in a series of works \cite{Gursoy:2007cb,Gursoy:2007er,Casero:2007ae,Iatrakis:2010zf, Iatrakis:2010jb,Jarvinen:2011qe,Alho:2012mh,Alho:2013hsa,Jarvinen:2015ofa}, where the properties of QCD thermodynamics and chiral phase transition in the Veneziano limit have been investigated in detail.

The temperature and chemical potential effects on the chiral condensate have been studied in the original soft-wall model \cite{Colangelo:2011sr} and also in the modified versions \cite{Bartz:2016ufc,Bartz:2017jku}, which cannot reproduce the basic structures of the QCD phase diagram. As the original one lacks spontaneous chiral symmetry breaking \cite{Karch:2006pv,Colangelo:2011sr}, we have proposed a simply improved soft-wall model with a quartic term of bulk scalar field and a running bulk scalar mass, which can give consistent light meson spectra and the correct chiral transition behaviors in the two-flavor case \cite{Fang:2016nfj}. The generalization to the $2+1$ flavor case has also been considered in \cite{Fang:2018vkp}, where the standard scenario of the quark-mass phase diagram (i.e., the Columbia plot) has been realized. Naturally, the further step is to study the chiral phase transition with finite chemical potential and the QCD phase diagram in the $\mu-T$ plane, which is just the theme of this work.

The paper is organized as follows. In Sec.\ref{model-intro}, we outline the improved soft-wall model with finite chemical potential, and the equation of motion (EOM) with proper boundary condition will be derived. The numerical results for the chiral phase transition at finite chemical potential will be given in Sec.\ref{numer-chiraltrans}, and the calculated chiral phase diagram will be presented in Sec.\ref{muTdiagram}, where the main results of our work will be given. In Sec.\ref{conclution}, we give a summary of our work and discuss some issues on the chiral phase transition in this framework.

\section{The improved soft-wall model with finite chemical potential}\label{model-intro}

\subsection{Bulk background}
To investigate the chemical potential effects on the chiral phase transition, we use the AdS/Reissner-Nordstrom (AdS/RN) black hole as the bulk background following Refs. \cite{Lee:2009bya,Colangelo:2011sr}. The AdS/RN black hole is the solution of the bulk gravity coupled to a $U(1)$ gauge field with the prescription $A_i=A_z=0$, $A_0=A_0(z)$ and the metric ansatz 
\begin{equation}\label{AdS-BH}
ds^2=e^{2A(z)}\left(f(z)dt^2-dx^{i\,2}-\frac{dz^2}{f(z)}\right), \quad A(z)=-\mathrm{log}\frac{z}{L},
\end{equation}
which gives
\begin{align}\label{AdS/RN-BH}
A_0(z) &= \mu -\kappa qz^2 ,      \nonumber \\
f(z) &=1-(1+Q^2)\left(\frac{z}{z_h}\right)^4 +Q^2\left(\frac{z}{z_h}\right)^6 , \quad Q=qz_h^3,
\end{align}
where $q$ is the charge of the black hole with horizon $z_h$, and $\kappa$ is a dimensionless constant which scales as $\sqrt{N_c}$ \cite{Lee:2009bya}. Below we will take $\kappa=1$ as in Ref. \cite{Colangelo:2011sr}. The chemical potential $\mu$ is determined by the boundary condition $A(z_h)=0$ as
\begin{align}\label{mu1}
\mu =\kappa\frac{Q}{z_h} =\frac{Q}{z_h} ,
\end{align}
and the Hawking temperature is given by
\begin{align}\label{T1}
T =\frac{1}{4\pi}\left|\frac{df}{dz}\right|_{z_{h}} =\frac{1}{\pi z_h}\left(1-\frac{Q^2}{2}\right), \quad  0<Q<\sqrt{2} .
\end{align}
Eliminating $Q$ from Eqs. (\ref{mu1}) and (\ref{T1}), we obtain the relation between $\mu$ and $T$
\begin{align}\label{T-mu-relat}
T =\frac{1}{\pi z_h}\left(1-\frac{\mu^2z_h^2}{2}\right), \quad  0<\mu z_h<\sqrt{2} .
\end{align}

\subsection{Model action}
As shown in our previous work \cite{Fang:2016nfj}, the improved soft-wall model with a $z$-dependent bulk scalar mass and a quartic scalar-coupling term can realize consistently the properties of linear confinement and spontaneous chiral symmetry breaking. In the $2+1$ flavor case, several studies have shown that the flavor-mixing effect introduced by a 't Hooft determinant term of the bulk scalar field is necessary for the correct realization of chiral phase transition \cite{Li:2016smq,Bartz:2017jku,Fang:2018vkp}. Our recent work shows that the improved soft-wall model with the determinant term added can reproduce the expected quark-mass dependence of the order of chiral phase transition \cite{Fang:2018vkp}. The action of the improved soft-wall model in the $2+1$ flavor case can be written as
\begin{equation}\label{2+1-act}
S =\int d^{5}x\,\sqrt{g}\,e^{-\Phi(z)}\left[\mathrm{Tr}\{|DX|^{2}-m_5^2(z)|X|^{2}
-\lambda |X|^{4}-\frac{1}{4g_{5}^2}(F_{L}^2+F_{R}^2)\} -\gamma\,\re\{\det X\}\right] ,
\end{equation}
where the covariant derivative of the bulk scalar field is $D^MX=\partial^MX-i A_L^MX+i X A_R^M$ and the chiral gauge field strength $F_{L,R}^{MN}=\partial^MA_{L,R}^N-\partial^NA_{L,R}^M-i[A_{L,R}^M,A_{L,R}^N]$. The dilaton field $\Phi(z)=\mu_g^2\,z^2$ is required for the property of linear confinement \cite{Karch:2006pv}. Theoretically, the running quark-mass anomalous dimension can induce a $z$-dependent bulk scalar mass $m_5(z)$ which, together with the quartic scalar-coupling term, are intimately related to the chiral dynamics \cite{Fang:2016cnt,Fang:2016nfj}. The UV limit of $m_5^2(z)$ is determined by the mass-dimension relation in AdS/CFT \cite{Erlich:2005qh}, while the IR limit of $m_5^2(z)$ can be well constrained by the mass split of the chiral partners. Below we will choose the simplest form $m_5^2(z)=-3-\mu_c^2\,z^2$ as in \cite{Fang:2016nfj}. 

\subsection{Equation of motion}
As we only consider the issue of chiral phase transition, the sectors related to the chiral gauge fields and the fluctuations of scalar and pseudoscalar modes can be safely neglected. Thus the bulk action relevant for our discussion is just
\begin{equation}\label{2+1-vev-act}
S_{\chi} =\int d^{5}x\,\sqrt{g}\,e^{-\Phi(z)}\left[\mathrm{Tr}\{\partial^{z}\VEV{X}\partial_{z}\VEV{X} -m_5^2(z)\VEV{X}^{2}
-\lambda\VEV{X}^{4}\} -\gamma\,\det\VEV{X}\right],
\end{equation}
where the vacuum expectation value (VEV) of the scalar field responsible for the chiral symmetry breaking takes the form
\begin{equation}\label{VEVs}
\langle X \rangle=\frac{1}{\sqrt{2}}
\begin{pmatrix}
\chi_u(z) & 0 & 0 \\
0 & \chi_d(z) & 0 \\
0 & 0 & \chi_s(z)
\end{pmatrix}.
\end{equation}
The EOM of the VEV $\VEV{X}$ can be obtained from the variation of the above action:
\begin{align}
& \chi_{u}'' +\left(\frac{f'(z)}{f(z)}+3A'(z)-\Phi'(z)\right)\chi'_{u} -\frac{e^{2A(z)}}{f(z)}\left(m_5^2(z)\chi_u +\lambda\chi_u^3 +\frac{\gamma}{2\sqrt{2}}\chi_u\chi_s \right) =0 \,, \label{vevX-eom1} \\
& \chi''_{s} +\left(\frac{f'(z)}{f(z)}+3A'(z)-\Phi'(z)\right)\chi'_{s} -\frac{e^{2A(z)}}{f(z)}\left(m_5^2(z)\chi_s +\lambda\chi_s^3 +\frac{\gamma}{2\sqrt{2}}\chi_{u}^{2}\right) =0 \,,  \label{vevX-eom2}
\end{align}
where the flavor-mixing terms come from the 't Hooft determinate term of the action (\ref{2+1-vev-act}). Note that we have taken $\chi_d=\chi_u$ for the $2+1$ flavor case.

\subsection{Boundary condition}
According to the AdS/CFT dictionary \cite{Erlich:2005qh}, the UV asymptotic forms of $\chi_u$ and $\chi_s$ can be obtained from Eqs. (\ref{vevX-eom1}) and (\ref{vevX-eom2}) as
\begin{align}
\chi_u(z \sim 0) &=m_u\,\zeta\,z-\frac{m_u\, m_s\,\gamma \, \zeta^2}{2\sqrt2}z^2+\frac{\sigma_u}{\zeta}z^3+\frac{1}{16}m_u\,\zeta\left(-m_s^2 \, \gamma^2\,\zeta^2-m_u^2\, \gamma^2 \,\zeta^2 \right.\nonumber\\
&\,\quad \left. +8 m_u^2 \,\zeta^2 \,\lambda+16 \,\mu_g^2-8 \,\mu_c^2\right)z^3 \,\mathrm{ln}z +\cdots \,, \label{asyExpOfChius1}  \\
\chi_s(z \sim 0) &=m_s\,\zeta\,z-\frac{m_u^2\,\gamma \, \zeta^2}{2\sqrt2}z^2+\frac{\sigma_s}{\zeta}z^3+\frac{1}{8}\left(-m_s\,m_u^2 \, \gamma^2\,\zeta^3+4\,m_s^3\,\zeta^3\,\lambda \right. \nonumber \\
&\,\quad \left. +8 m_s \,\zeta \,\mu_g^2-4 \,m_s\,\zeta\,\mu_c^2 \right)z^3 \,\mathrm{ln}z +\cdots \,, \label{asyExpOfChius2}
\end{align}
where $m_{u,s}$ denote the quark mass and $\sigma_{u,s}$ the chiral condensates of u,s quarks, and the normalization $\zeta=\frac{\sqrt{N_c}}{2\pi}$ is fixed by the large-$N_c$ scaling analysis of the quark mass and chiral condensate \cite{Cherman:2008eh}. The aim of this work is just to study the temperature and chemical potential effects on the chiral condensates.

The IR boundary condition is determined by the regular condition of the VEV $\chi_{u,s}(z)$ at the black-hole horizon $z=z_h$ \cite{Fang:2018vkp}, which yields
\begin{align}
\frac{12\,\chi_u(z_h) +4\,z_h^2\,\mu_c^2\,\chi_u(z_h) -\sqrt{2}\,\gamma\,\chi_s(z_h)\,\chi_u(z_h) -4\,\lambda\,\chi_u^3(z_h)}{8(Q^2 -2)z_h} +\chi_u'(z_h) &=0, \label{boundCond2-1} \\
\frac{12\,\chi_s(z_h) +4\,z_h^2\,\mu_c^2\,\chi_s(z_h) -\sqrt{2}\,\gamma\,\chi_u^2(z_h) -4\,\lambda\,\chi_s^3(z_h)}{8(Q^2 -2)z_h} +\chi_s'(z_h) &=0. \label{boundCond2-2}
\end{align}
To obtain the chiral transition behaviors with finite temperature and chemical potential, we need to solve the boundary-value problem of the coupled nonlinear ordinary differential Eqs. (\ref{vevX-eom1}) and (\ref{vevX-eom2}). One can refer to Ref. \cite{Fang:2018vkp} for the details of the numerical method we used.

\section{Chiral phase transition with finite chemical potential}\label{numer-chiraltrans}

\subsection{Parameter choice and the $\mu=0$ case}

\begin{figure}
\centering
\includegraphics[width=75mm,clip=true,keepaspectratio=true]{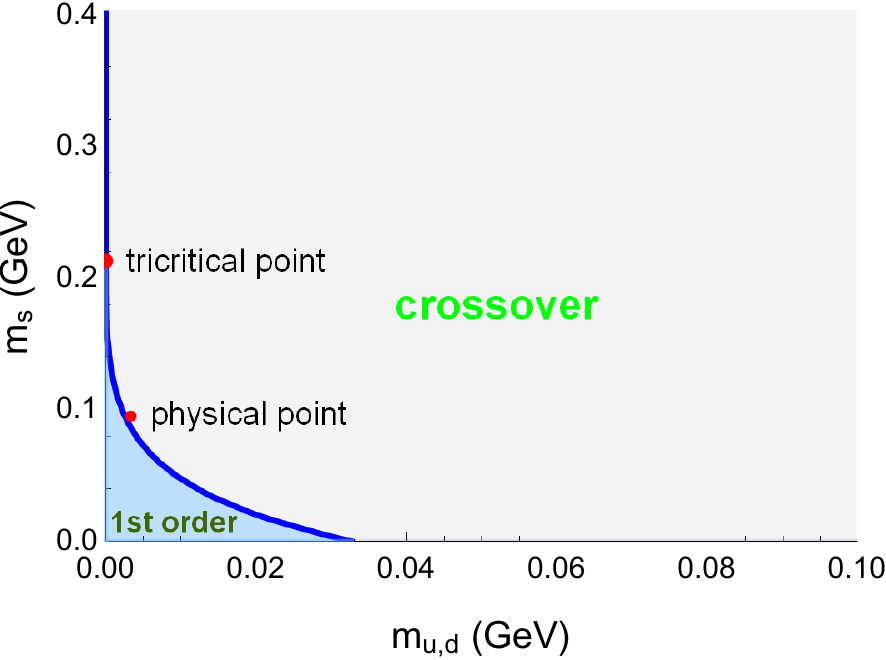}
\caption{The calculated phase diagram in the quark-mass plane from the improved soft-wall model with $\mu=0$ \cite{Fang:2018vkp}. The blue curve represents the second-order critical line.} 
\label{ads-phase-diagram1}
\end{figure}

There are four parameters in the improved soft-wall model, i.e., the scalar-coupling constants $\lambda$, $\gamma$ and the parameters $\mu_g$, $\mu_c$ in the dilaton and the running bulk scalar mass. Previous studies have shown that the value of $\mu_g$ is related to the linear Regge behavior of the highly-excited hadron spectrum, while $\lambda$, $\gamma$ and $\mu_c$ are crucial for the property of spontaneous chiral symmetry breaking and determine the correct behaviors of chiral phase transition \cite{Fang:2016nfj,Fang:2018vkp}. As in \cite{Fang:2018vkp}, we choose the parameter values to be $\gamma=-22.6$, $\lambda=16.8$, $\mu_{c}=1180\,\mathrm{MeV}$ and $\mu_{g}=440\,\mathrm{MeV}$, which can yield the standard scenario of the phase diagram in the quark-mass plane with a crossover transition at the physical point ($m_{u,d}=3.5\,\mathrm{MeV}$ and $m_{s}=96\,\mathrm{MeV}$ \cite{Agashe:2014kda}). The main results of Ref.\cite{Fang:2018vkp} are presented in Fig.\ref{ads-phase-diagram1}, where the second-order critical line separates the first-order phase transition at small quark-mass region from the crossover transition at large quark-mass region, and a tricritical point on the $m_{u,d}=0$ boundary can be naturally obtained in this model.

\subsection{The finite $\mu$ case with zero quark mass}\label{sec-zeromass-case}

Before considering the chiral transition behavior at physical quark mass, we first address the simplest case with $m_{u,d} =m_{s} =0$, in which the bulk scalar VEV $\chi_u$ and $\chi_s$ are equal with each other, and we define $\chi\equiv\chi_{u}=\chi_{s}$. The EOM then reduces to a single one in this case
\begin{align}\label{zeromass-vevX-eom1}
\chi'' +\left(\frac{f'}{f}+3A'-\Phi'\right)\chi' -\frac{e^{2A}}{f}\left(m_5^2\chi +\lambda\chi^3 +\frac{\gamma}{2\sqrt{2}}\chi^2\right) &=0,
\end{align}
which has been considered in detail in the $\mu=0$ case \cite{Chelabi:2015gpc}. Generally, Eq. (\ref{zeromass-vevX-eom1}) admits two solutions, i.e., the trivial one $\chi=0$ with zero chiral condensate and $\chi\neq 0$ with nonzero condensate.

\begin{figure}
\begin{center}
\includegraphics[width=68mm,clip=true,keepaspectratio=true]{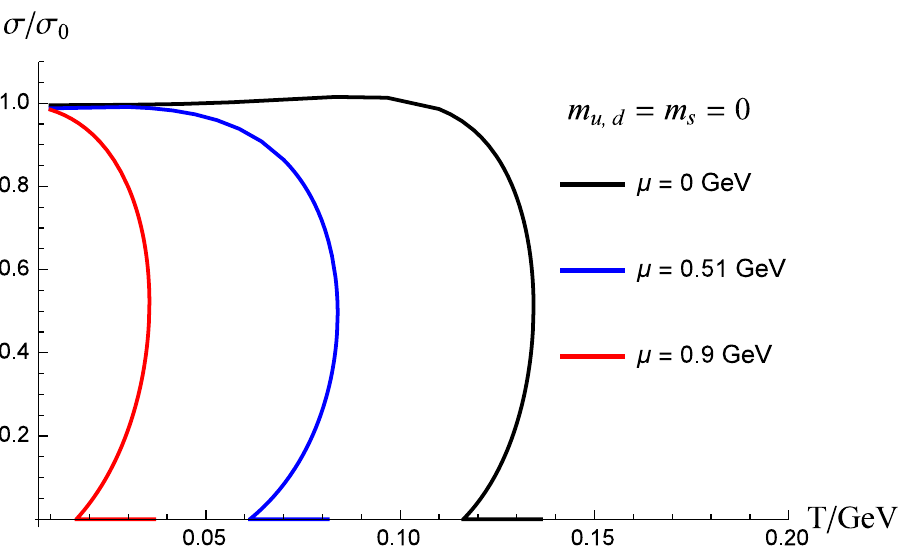}
\hspace*{1cm}
\includegraphics[width=68mm,clip=true,keepaspectratio=true]{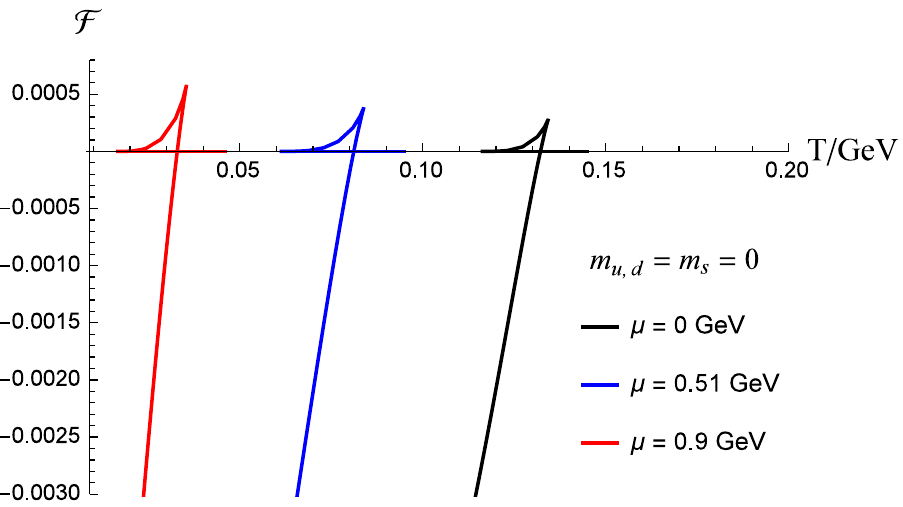} \vskip -1cm \hskip 5.5 cm
\end{center}
\caption{Left: chiral transition behavior of the rescaled condensate $\frac{\sigma}{\sigma_0}$ with temperature $T$ at chemical potentials $\mu=0, 0.51, 0.9\,\mathrm{GeV}$ in the $m_{u,d} =m_{s} =0$ case; Right: the corresponding free energies $\mathcal{F}(T)$ of the bulk scalar VEV at $\mu=0, 0.51, 0.9\,\mathrm{GeV}$.}
\label{sigma-T-mu-zeromass}
\end{figure}

We present the numerical results for the chiral phase transition at three different chemical potentials $\mu=0, 0.51, 0.9\,\mathrm{GeV}$ in Fig.\ref{sigma-T-mu-zeromass} (left panel), where the chiral condensate $\sigma$ is rescaled by the zero-temperature value $\sigma_0$. We can see obviously that the chiral phase transition in the $m_{q} =0$ case is a first-order one with a decreasing $T_c$ as $\mu$ increases. At each $T_c$, the value of $\sigma$ drops to zero with the chiral symmetry restored. To pin down the value of $T_c$ at each $\mu$, we can compare the free energy of the $\chi=0$ solution with that of the $\chi\neq 0$ solution, which can be derived from the bulk action of $\VEV{X}$ (see Appendix \ref{freeenergy}). The numerical results of the free energies at $\mu=0, 0.51, 0.9\,\mathrm{GeV}$ are also shown in Fig.\ref{sigma-T-mu-zeromass} (right panel), where the self intersection of each free-energy curve determines the critical temperature $T_c$.

\subsection{The finite $\mu$ case with physical quark mass}

\begin{figure}
\begin{center}
\includegraphics[width=68mm,clip=true,keepaspectratio=true]{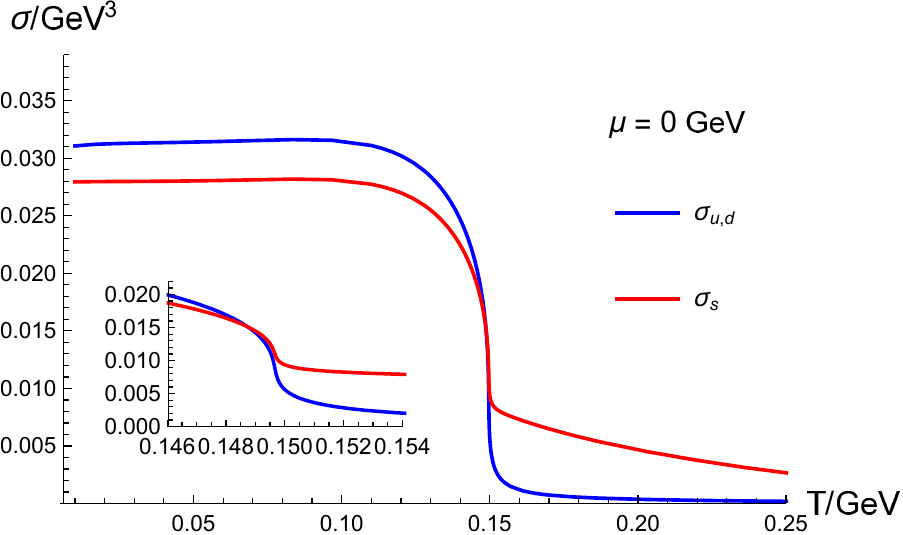}
\hspace*{0.6cm}
\includegraphics[width=68mm,clip=true,keepaspectratio=true]{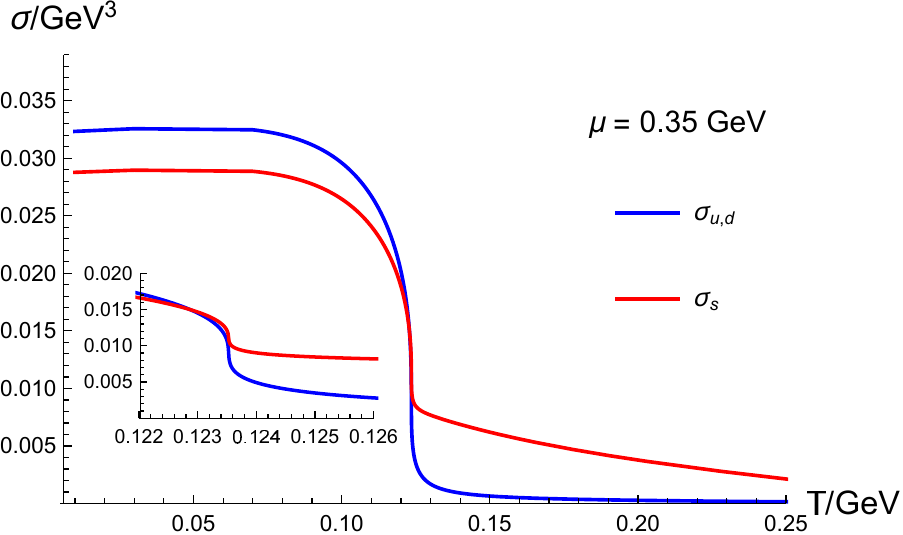}
\vspace{0.35cm} \\ 
\includegraphics[width=68mm,clip=true,keepaspectratio=true]{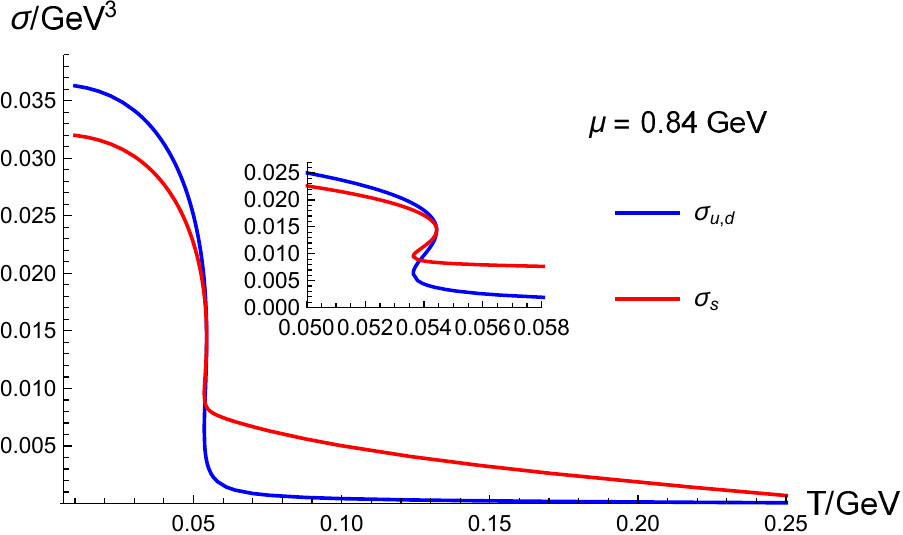}
\hspace*{0.6cm}
\includegraphics[width=68mm,clip=true,keepaspectratio=true]{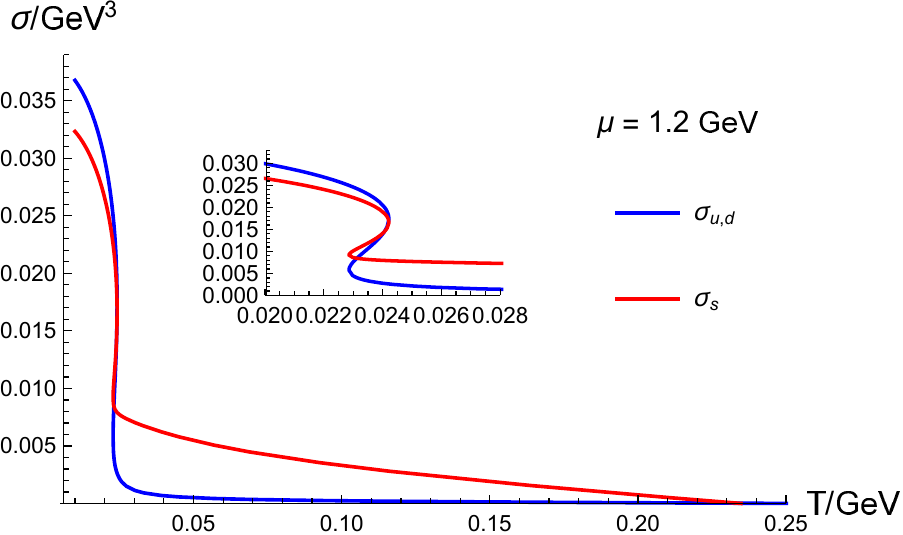}
\vskip -1cm \hskip 0.7 cm
\end{center}
\caption{Chiral transition behaviors of the condensates $\sigma_{u,d}$ and $\sigma_s$ with temperature $T$ at chemical potentials $\mu=0, 0.35, 0.84, 1.2\,\mathrm{GeV}$ in the case of physical quark mass ($m_{u,d}=3.5\,\mathrm{MeV}$, $m_{s}=96\,\mathrm{MeV}$).}
\label{sigma-T-mu-physicalmass}
\end{figure}

At the physical point, we need to solve the two coupled Eqs. (\ref{vevX-eom1}) and (\ref{vevX-eom2}) numerically to obtain the values of $\sigma_u$ and $\sigma_s$, which show different chiral transition behaviors due to the mass difference of $u, s$ quarks. However, the chiral transitions of $\sigma_u$ and $\sigma_s$ are always entangled with each other in our model, i.e., they have the same (pseudo-)critical temperature. At $\mu=0$, we have known that the physical point has a crossover transition with $T_c\simeq 150\,\mathrm{MeV}$ \cite{Fang:2018vkp}. Here we present the numerical results for the chiral transitions of $\sigma_u$ and $\sigma_s$ at four different chemical potentials $\mu=0, 0.35, 0.84, 1.2\,\mathrm{GeV}$ in Fig.\ref{sigma-T-mu-physicalmass}, from which we find that the crossover transition at $\mu=0$ will turn into a first-order one at large enough chemical potentials, and a second-order phase transition appears at some intermediate chemical potential ($\mu_c\simeq 0.35\,\mathrm{GeV}$ in our case), which signifies the existence of a CEP, as expected from the general considerations of the $\mu-T$ phase diagram. The chiral transition temperature $T_c$ decreases towards zero as $\mu$ increases, which is also consistent with expectations.

\begin{figure}
\begin{center}
\includegraphics[width=68mm,clip=true,keepaspectratio=true]{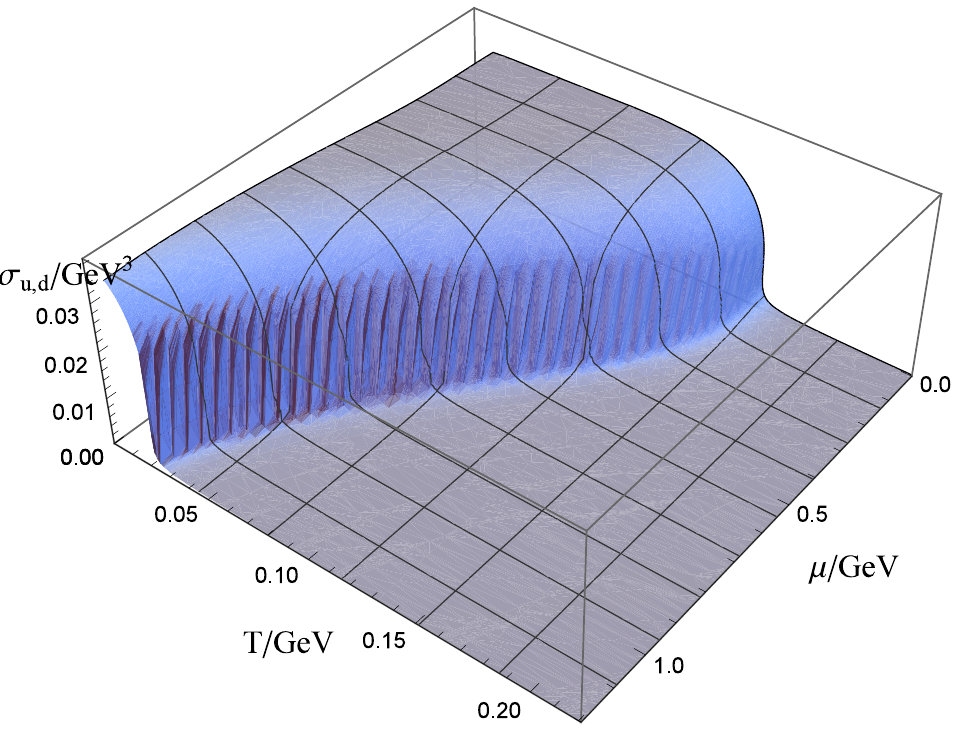}
\hspace*{1cm}
\includegraphics[width=68mm,clip=true,keepaspectratio=true]{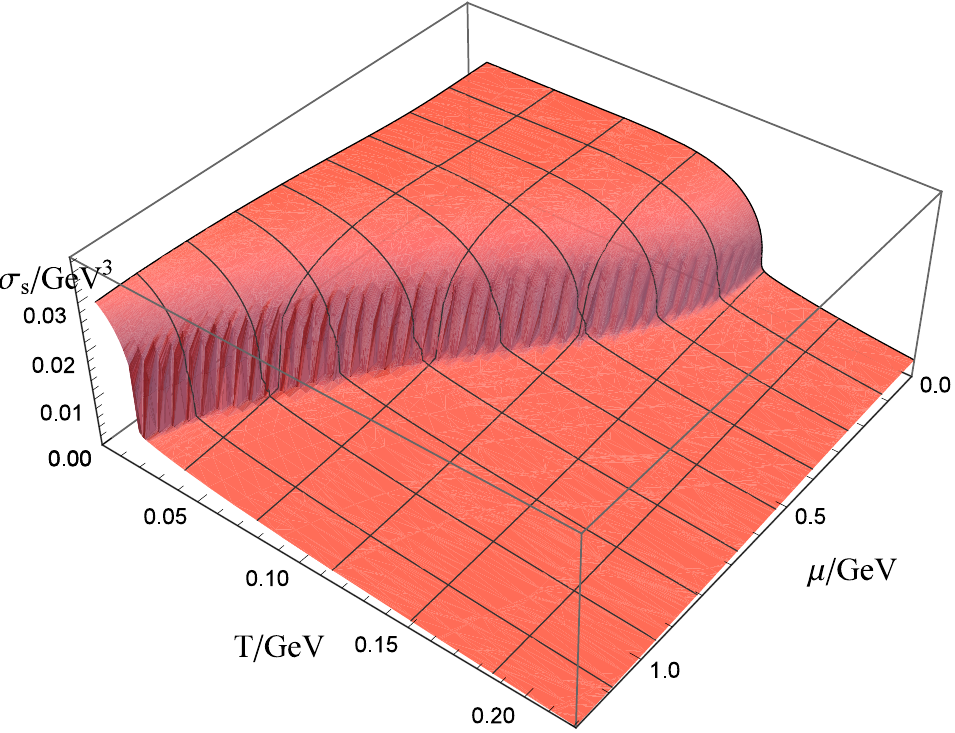} \vskip -1cm \hskip 5.5 cm
\end{center}
\caption{Chiral condensates $\sigma_{u,d}$ (left) and $\sigma_s$ (right) as functions of temperature $T$ and chemical potential $\mu$ in the case of physical quark mass ($m_{u,d}=3.5\,\mathrm{MeV}$, $m_{s}=96\,\mathrm{MeV}$).}
\label{sigma-T-mu-3D}
\end{figure}

To give a detailed characterization for the chiral transitions of $\sigma_u$ and $\sigma_s$ at finite $\mu$, we show the 3D plot of the chiral condensate $\sigma_{u,s}$ as a function of $T$ and $\mu$ in Fig.\ref{sigma-T-mu-3D}, where the plateau region with larger condensates and the region with smaller ones are divided by a critical transition line in between. The decease of transition temperature with increasing $\mu$ can be seen obviously. We also see that the chiral condensates $\sigma_{u,s}$ undergo a similar transition with the chemical potential $\mu$ as that with the temperature $T$.

\section{Chiral phase diagram}\label{muTdiagram}

The main concern for the $\mu-T$ phase diagram in our case is about the existence and location of the CEP. As far as we know, there are few works addressing this issue from the angle of chiral phase transition in AdS/QCD. The original soft-wall model is too simple to realize the correct chiral transition behavior \cite{Colangelo:2011sr}. One can refer to \cite{Bartz:2016ufc,Bartz:2017jku} for similar considerations in another modified soft-wall model, where no CEP shows in their calcultion. From the above study, we see that the chiral phase transition at physical quark mass in our improved soft-wall model contains naturally a CEP linking the crossover transition at smaller $\mu$ with the first-order phase transition at larger $\mu$.

\begin{figure}
\begin{center}
\includegraphics[width=68mm,clip=true,keepaspectratio=true]{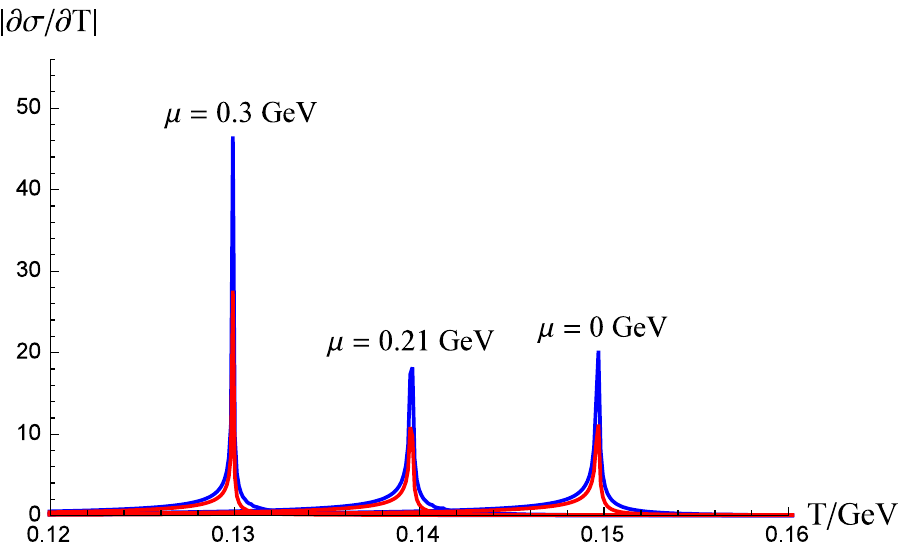}
\hspace*{1cm}
\includegraphics[width=68mm,clip=true,keepaspectratio=true]{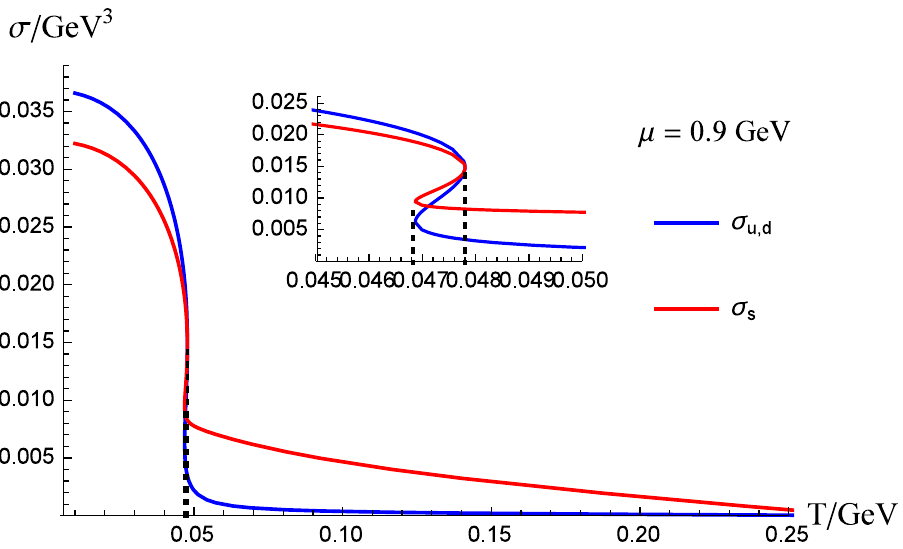} \vskip -1cm \hskip 5.5 cm
\end{center}
\caption{Determination of the critical temperature $T_c$ for the crossover (left) and first-order (right) transitions in the case of physical quark mass. Left: the first derivative of chiral condensate $\big|\frac{\partial\sigma_q}{\partial T}\big|$ at $\mu=0, 0.21, 0.3\,\mathrm{GeV}$; Right: the first-order phase transition at $\mu=0.9\,\mathrm{GeV}$ with $T_c$ restricted to the region between the two dashed vertical lines.}
\label{critical-T}
\end{figure}

To obtain the $\mu-T$ phase diagram, we need to fix the transition temperature $T_c$ at each $\mu$. In the $m_q=0$ case, $T_c$ can be determined by the stability criteria of free energy (see appendix \ref{freeenergy}), while in the case of physical quark mass, we need to fix $T_c$ for the crossover and first-order transitions separately. The crossover transition temperature can be defined as the minimum of the first derivative of chiral condensate $\frac{\partial\sigma_q}{\partial T}$, which is shown in Fig.\ref{critical-T} (left panel) for $\mu=0, 0.21, 0.3\,\mathrm{GeV}$. For the first-order phase transition with $\mu>0.35\,\mathrm{GeV}$, we only restrict the critical temperature to the region between the two inflections of the chiral transition (see the right panel of Fig.\ref{critical-T} for the $\mu=0.9\,\mathrm{GeV}$ case).

\begin{figure}
\centering
\includegraphics[width=75mm,clip=true,keepaspectratio=true]{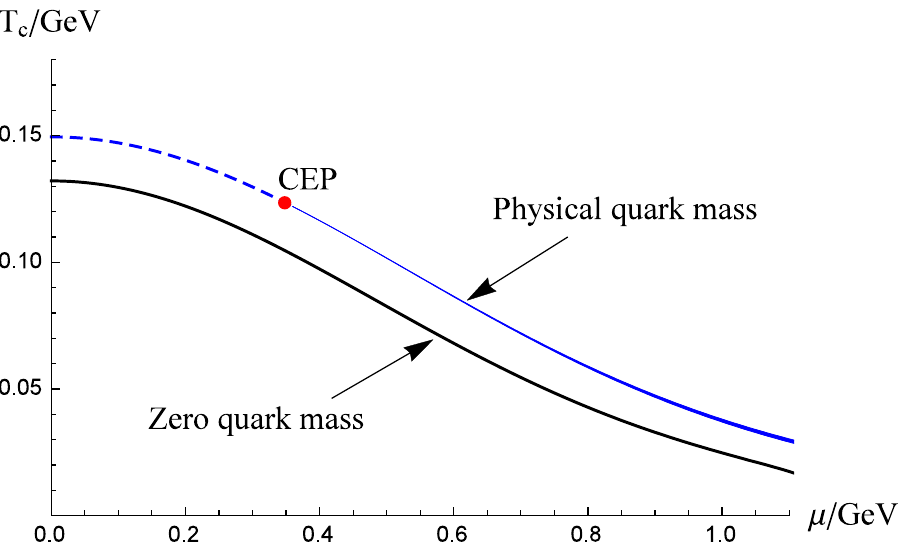}
\caption{The $\mu-T$ phase diagram in the case of zero quark mass ($m_{u,d} =m_{s} =0\,\mathrm{MeV}$) and physical quark mass ($m_{u,d}=3.5\,\mathrm{MeV}$, $m_{s}=96\,\mathrm{MeV}$).}
\label{mu-T-diagram}
\end{figure}

The $\mu-T$ phase diagram obtained from our improved soft-wall model is shown in Fig.\ref{mu-T-diagram}, where $T_c$ as a function of $\mu$ has been given for both zero quark mass (black curve) and physical quark mass (blue curve), from which one can see that the transition temperature decreases with increasing chemical potential. There is only first-order phase transition in the case of zero quark mass, as can be seen from Sec.\ref{sec-zeromass-case}. The interesting point comes from the case of physical quark mass, in which the crossover transition turns into a first-order one with the increase of $\mu$ and a CEP naturally appears at $\mu_c\simeq0.35\,\mathrm{GeV}$, which is consistent with the current perspective on the QCD phase diagram. Note that the specific value of the CEP is not our concern in this work as the parameter choice in our calculation is more or less arbitrary without any constraints from other physical quantities. The transition temperature is only calculated up to $\mu=1.1\,\mathrm{GeV}$. We keep cautious on our results at large chemical potentials and low temperatures, which might beyond the scope of this simple AdS/QCD model for the description of QCD phase transition \cite{Rougemont:2015wca}.

\begin{figure}
\centering
\includegraphics[width=85mm,clip=true,keepaspectratio=true]{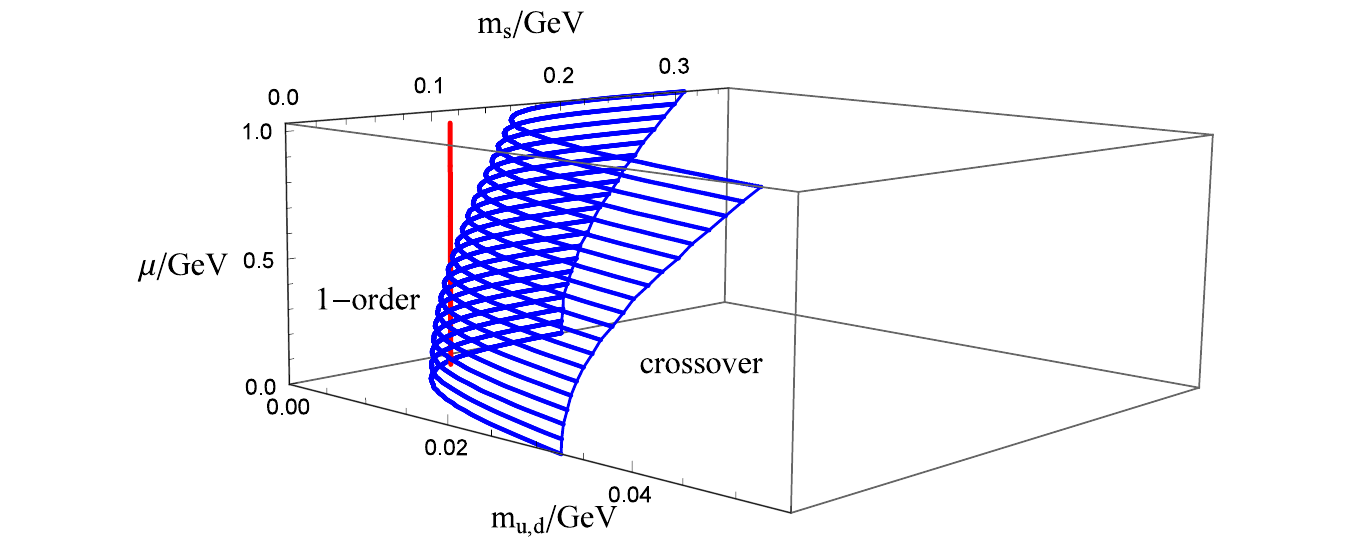}
\caption{The calculated critical surface (blue) separating first-order phase transition from crossover transition. The red vertical line intersecting with the critical surface corresponds to the physical point of quark mass.} 
\label{critical-surface}
\end{figure}

In Ref.\cite{Fang:2018vkp}, we have studied the flavor and quark-mass dependence of the chiral phase transition in the improved soft-wall model, and reproduce the standard phase diagram in the quark-mass plane, as shown in Fig.\ref{ads-phase-diagram1}, where the second-order critical line separates the crossover region from the region of first-order phase transition. Here we also investigate the change of the second-order line with different chemical potentials. The numerical results are shown in Fig.\ref{critical-surface}, where the second-order lines constitute a critical surface, which is crossed through by a red vertical line locating at the physical point.

\section{Conclusion and discussion}\label{conclution}

In this work, we studied the chemical potential effects on chiral phase transition in an improved soft-wall model, and obtained the $\mu-T$ phase diagram at both zero quark mass and physical quark mass. In the case of physical quark mass, the crossover transition turns into a first-order one with the increase of chemical potential, and a CEP appears at $\mu_c\simeq0.35\,\mathrm{GeV}$. These features of the chiral phase transition obtained from our model agree with the current picture of the QCD phase diagram.

Some cautions on our work should be given here. To introduce the temperature and chemical potential, we use the AdS/RN black-hole solution as the fixed background without consideration of the flavor back-reaction, which should be studied further to make sure that the qualitative behaviors of chiral phase transition presented here would not be changed. One main suspicion for the relevance of our results as well as many other works to QCD phase transition is that we only used the black hole solution as the thermal background. However, it is generally believed that the gravity dual of deconfining phase transition has a Hawking-Page type with the AdS black hole solution only corresponding to the deconfined phase, while the confined phase at low temperature is characterized by a thermal AdS solution \cite{Herzog:2006ra}. Although the large $N_c$ analyses from (super-)Yang-Mills theories support this argument, the subtlety is that such a Hawking-Page phase transition relies sensitively on the bulk geometry and the profile of the dilaton field solved from the coupled gravity-dilaton system \cite{Herzog:2006ra}. Further more, there would be a discontinuity in the chiral phase transition if one admits two different bulk gravity solutions for the low and high temperature phase, and this is incompatible with the smooth crossover transition at physical quark mass, at least in the framework of soft-wall models \cite{Colangelo:2011sr}. How to reconcile these theoretical analyses with the phenomenological studies is a pressing issue for the current research of AdS/QCD.

There have been many holographic studies on QCD phase transition and QCD thermodynamics with the single AdS black-hole solution as the bulk background (for some early ones, refer to \cite{Andreev:2006eh,Andreev:2009zk,Gubser:2008yx,Gubser:2008ny,Gubser:2008sz}). As in \cite{Gubser:2008yx,Gubser:2008ny}, we may view the AdS black-hole description as an approximation which loses its validity gradually as the temperature decreases below $T_c$. Setting these issues aside, our work shows for the first time that a simply improved soft-wall model can produce rich phase structure similar as that of QCD from the analysis of chiral phase transition at finite chemical potential. Thus we hope that it may serve as a useful guidance for the further development of AdS/QCD in both theoretical and phenomenological researches.

In the study of the chemical potential effects on chiral phase transition, we only focus on the qualitative features of the phase diagram with no intention to give a quantitative description. To extract the value of CEP which can be compared with lattice or other model results, we need to fix the parameters of the improved soft-wall model by other physical quantities, such as the hadron spectrum. The dynamics of the bulk background can be introduced through a Maxwell-Einstein-dilaton system. Then the deconfining and chiral phase transitions can be considered simultaneously, and the relevant issues such as the relation between these two phase transitions of QCD can be addressed in this framework of AdS/QCD.

\section*{Acknowledgements}
This work was supported by National Science Foundation of China (NSFC) (11690022, 11475237, 11121064) and Strategic Priority Research Program of the Chinese Academy of Sciences (XDB23030100) as well as the CAS Center for Excellence in Particle Physics (CCEPP), and also supported by the Fundamental Research Funds of the Central Universities (No.531107051199).

\appendix
\section{Free energy of the scalar VEV}\label{freeenergy}

In this work, we use the stability criteria of free energy to determine the critical temperature $T_c$ of the first-order phase transition in the $m_q=0$ case. We would like to derive the formula of the free energy in our case and give some discussions. According to the AdS/CFT correspondence, we have the equivalence of the partition functions $Z_{\text{gauge}}=Z_{\text{gravity}}\simeq e^{-S_E}$, which leads to the relation between the free energy of the thermodynamic system in the boundary and the on-shell Euclidean action of the bulk gravity: $F=S_E$. Substituting Eq. (\ref{VEVs}) into Eq. (\ref{2+1-vev-act}) and using Eqs. (\ref{vevX-eom1}) and (\ref{vevX-eom2}), we can obtain the expression for the free energy of the bulk scalar VEV:
\begin{align}\label{2+1freeenergy}
\mathcal{F} &\equiv \frac{S_E}{V^3} =-\int_{\epsilon}^{z_h} dz\,\sqrt{g}\,e^{-\Phi}\left[\mathrm{Tr}\{g^{zz}(\partial_{z}\VEV{X})^2 -m_5^2\VEV{X}^{2}
-\lambda\VEV{X}^{4}\} -\gamma\,\det\VEV{X}\right]     \nonumber\\
&=-\sum_{q=u,d,s}\frac{1}{2}\left(e^{3 A-\Phi}f\chi_q\chi_q'\right)\big|_{\epsilon}
-\int_{\epsilon}^{z_h}\mathrm{d}z e^{5A-\Phi}\left[\frac{\gamma}{4\sqrt{2}}\chi_u\chi_d\chi_s +\frac{\lambda}{4}\left(\chi_u^4 +\chi_d^4 +\chi_s^4\right)\right],
\end{align}
where the divergent volume integration of the three-dimensional space $V^3$ has been removed, and the ultraviolet cutoff $z=\epsilon$ has been used to regularize the possible divergence of the bulk action.

At zero quark mass, the boundary term in the free-energy formula vanishes and the integration term is finite as $\epsilon\to 0$, thus we can safely use the above formula to calculate the free energy of the scalar VEV solution. For the case of physical quark mass, both the boundary term and the integration term are divergent. In principle, one can separate the finite part from the divergent one which is independent of $\mu$ and $T$, and then add counter terms to cancel the divergent part. However, the calculated free energy following this procedure exhibits discontinuity in the region with rapid chiral transition, which seems to contradict with the general principles of thermodynamics. There may be other ingredients missed in our consideration of the free-energy calculation at physical quark mass. Hence we are content in this work to just constrain the critical temperature to a small range, which has been illustrated in Fig.\ref{critical-T}.

\bibliography{refs-AdSQCD}
\end{document}